\documentclass[
  ,final            
  ,numberedheadings 
  ]
  {aipproc}

\layoutstyle{6x9}

\usepackage{times}
\usepackage{amssymb}
\usepackage{amsmath}
\usepackage{amscd}
\usepackage{latexsym}
\usepackage{bbm}
\usepackage{epsfig}

\def\a{\alpha}
\def\b{\beta}
\def\g{\gamma}
\def\de{\delta}
\def\eps{\epsilon}
\def\ve{\varepsilon}

\def\h{\eta}
\def\th{\theta}
\def\vt{\vartheta}

\def\la{\lambda}
\def\m{\mu}
\def\n{\nu}
\def\r{\rho}
\def\s{\sigma}
\def\p{\phi}
\def\vp{\varphi}
\def\Ga{\Gamma}

\def\sfrac#1#2{{\textstyle\frac{#1}{#2}}}
\def\rd#1{\buildrel{_{_{\hskip 0.01in}\rightarrow}}\over{#1}}
\def\ld#1{\buildrel{_{_{\hskip 0.01in}\leftarrow}}\over{#1}}
\def\+{\dagger}
\def\={\ =\ }
\def\pa{\partial}
\def\pab{{\bar{\partial}}}
\def\>{\rangle}
\def\<{\langle}
\def\we{\wedge}

\newcommand{\unity}{\mathbbm{1}}

\newcommand{\e}{\,\mathrm{e}\,}
\newcommand{\im}{\,\mathrm{i}\,}
\newcommand{\diff}{\mathrm{d}}
\newcommand{\tr}{\mathrm{tr}}

\newcommand{\ch}{\mathrm{ch}}

\newcommand{\tah}{\mathrm{th}}
\newcommand{\vel}{\mathrm{v}}
\newcommand{\es}{\,\mathrm{e}_{{}_{\star}}\!\!}

\newcommand{\R}{{\mathbb{R}}}
\newcommand{\C}{{\mathbb{C}}}

\newcommand{\cE}{{\mathcal E}}
\newcommand{\Hcal}{{\cal H}}
\newcommand{\Acal}{{\cal A}}

\newcommand{\Ncal}{{\cal N}}

\newcommand{\Fcal}{{\cal F}}

\newcommand{\be}{\begin{equation}}
\newcommand{\ee}{\end{equation}}
\newcommand{\bea}{\begin{eqnarray}}
\newcommand{\eea}{\end{eqnarray}}
\newcommand{\bal}{\begin{aligned}}
\newcommand{\eal}{\end{aligned}}

\newcommand{\und}{\qquad{\text{and}}\qquad}

\newcommand{\zb}{{\bar{z}}}
\newcommand{\mb}{\bar{\mu}}
\newcommand{\At}{\widetilde{A}}
\newcommand{\Pt}{\widetilde{P}}
\newcommand{\Tt}{\widetilde{T}}

\newcommand{\ph}{\widehat{\phi}}
\newcommand{\Ph}{\widehat{\Phi}}
\newcommand{\tU}{{\textrm{U}}}

\newcommand{\sta}{\star}
\newcommand{\ar}{\alpha}

\newcommand{\bt}{\beta}

\newcommand{\se}{\sigma_1}

\newcommand{\sd}{\sigma_3}

\begin{document}

\begin{flushright}
CERN-PH-TH/2007-184\\
ITP--UH--22/07\\
\end{flushright}

\title{Noncommutative Solitons\footnote{
Talk given at the Third Mexican Meeting 
on Mathematical and Experimental Physics 
at El Colegio Nacional, Mexico City, 10-14 September 2007}}


\author{Olaf Lechtenfeld}{
  address={Theory Division, Physics Department, 
           CERN, CH-1211 Geneva 23, Switzerland},
  altaddress={on leave from: 
           Institut f\"ur Theoretische Physik, Leibniz Universit\"at Hannover}
}

\begin{abstract}
Solitonic objects play a central role in gauge and string theory 
(as, e.g., monopoles, black holes, D-branes, etc.). Certain string backgrounds
produce a noncommutative deformation of the low-energy effective field theory,
which allows for new types of solitonic solutions. 
I present the construction, moduli spaces and dynamics of Moyal-deformed 
solitons, exemplified in the 2+1 dimensional Yang-Mills-Higgs theory and its 
Bogomolny system, which is gauge-fixed to an integrable chiral sigma model 
(the Ward model). Noncommutative solitons for various 1+1 dimensional 
integrable systems (such as sine-Gordon) easily follow by dimensional and 
algebraic reduction. Supersymmetric extensions exist as well and are related 
to twistor string theory.
\end{abstract}

\maketitle


\section{Beating Derrick's theorem}

\subsection{Solitons in d=1+2 scalar field theory}

The simplest relativistic field theory in flat three-dimensional spacetime 
deals with a real scalar field~$\phi$ depending on time~$t$ and spatial
coordinates~$z=x{+}\im y$, governed by an action
\be
S^{(0)}[\phi] \= \int\!\diff{t}\,\diff{x}\,\diff{y}\; \bigl[
\sfrac12 \dot\phi^2 - \sfrac12 (\vec\nabla\phi)^2 - V(\phi) \bigr]\ ,
\ee
where I take the potential to be polynomial and bounded by zero from below, i.e.
\be
V'(\phi) \= v\prod_i (\phi-\phi_i) \und V(\phi)\ge 0 
\qquad\textrm{with}\quad V(\phi_0)=0\ , 
\ee
so that $\phi_i$ denote the (constant) extrema of~$V$, 
including the vacuum~$\phi_0$.
Static classical field configurations~$\ph$ extemize the energy
\be
E^{(0)}[\phi]\= \int\!\diff{x}\,\diff{y}\; \bigl[
\sfrac12 (\vec\nabla\phi)^2 + V(\phi) \bigr]\ ,
\ee
but Derrick's theorem rules out any interesting solutions:
Given some extremum $\ph(x,y)$ of~$E^{(0)}$, I may consider the family
$\ph_\la(x,y):=\ph(\frac{x}{\la},\frac{y}{\la})$ of rescaled configurations,
whose energy must be extremized by $\ph_1(x,y)$. 
However, the $\la$-variation of
\be
E^{(0)}(\la)\ :=\ E^{(0)}[\ph_\la] \=
\la^0 \!\int\!\diff{x}\,\diff{y}\;\sfrac12(\vec\nabla\ph)^2 \ +\
\la^2 \!\int\!\diff{x}\,\diff{y}\;V(\ph)
\ee
at $\la{=}1$ only vanishes for $V(\ph)\equiv0$, meaning $\ph=\phi_0$.
Thus, barring possible vacuum degeneracy, the only non-singular static
solution to the equation of motion is the vacuum.

\subsection{Noncommutative deformation}

One possible way out is a noncommutative deformation of the $xy$~plane,
the simplest of which is the Moyal deformation
\be\label{ncdef}
(x,y)\ \longrightarrow\ (X,Y)\qquad\textrm{with}\qquad
[X\,,\,Y]=\im\th=\textrm{const}\ .
\ee
For definiteness, one may realize the noncommuting spatial coordinates
$X$ and~$Y$ by infinite-dimensional matrices
\be
X = \frac{\sqrt{2\th}}{2} \left( \begin{smallmatrix}
0 & \sqrt{1} & & & \\ 
  \sqrt{1} & 0 & \sqrt{2} & & \\
  & \sqrt{2} & 0 & \sqrt{3} & \\[-7pt] 
  & & \sqrt{3} & 0 & \ddots \\[-7pt]
  & & & \ddots & \ddots \end{smallmatrix} \right)  \und
 Y = \frac{\sqrt{2\th}}{2\im} \left( \begin{smallmatrix}
0 & \sqrt{1} & & & \\ 
  \!\!\!{-}\sqrt{1} & 0 & \sqrt{2} & & \\
  & \!\!\!{-}\sqrt{2} & 0 & \sqrt{3} & \\[-7pt] 
  & & \!\!\!{-}\sqrt{3} & 0 & \ddots \\[-7pt]
  & & & \ddots & \ddots \end{smallmatrix} \right) \ ,
\ee
which act on some auxiliary (Fock) space~$\Fcal$.
More generally, sufficiently nice functions~$f(t,x,y)$ 
turn into bounded operators $F(t)$ via
\be\label{MWmap}
f(t,x,y)\quad\longrightarrow\quad
F(t) \= \textrm{Weyl-order}\ \bigl[ f(t,X,Y) \bigr] \qquad
\textrm{acting on $\Fcal$}\ ,
\ee
where every monomoial on the right-hand side is meant to be Weyl
(or symmetrically) ordered in $X$ and~$Y$.
Furthermore, derivatives become inner derivations in the operator algebra,
and the spatial integral translates to a trace over~$\Fcal$,
\be
\pa_x\ \to\  \sfrac{\im}{\th}[Y,\cdot\,] \ ,\quad
\pa_y\ \to\ -\sfrac{\im}{\th}[X,\cdot\,] \und
\int\!\diff{x}\,\diff{y}\,f = \pi\th\;\tr_\Fcal\,F \ .
\ee
I try to use corresponding small and capital letters for 
functions and operators, respectively.
The energy on the Moyal plane takes the form
\be\label{ncenergy}
E^{(\th)} \= \pi\,\tr\, \bigl\{
-\sfrac12[\sfrac{X}{\sqrt{\th}},\Phi]^2-\sfrac12[\sfrac{Y}{\sqrt{\th}},\Phi]^2
\ +\ \th\,V(\Phi) \bigr\}
\quad\buildrel{\th\to\infty}\over\longrightarrow\quad
\pi\th\;\tr\,V(\Phi)\ ,
\ee
which in the large-$\th$ limit is stationary for operators $\Ph$ subject to
\be\label{nceom}
0 \= V'(\Ph)\ \equiv\ v\,(\Ph{-}\phi_0)\,(\Ph{-}\phi_1)\cdots(\Ph{-}\phi_n)\ .
\ee

\subsection{New classical solutions}

Because of its operator-valuedness, Derrick's theorem is avoided, and
(\ref{nceom}) admits a large set of non-constant solutions
\cite{Gopakumar:2000zd},
\be
\Ph \= \sum_i \phi_i\,P_i \qquad\textrm{with}\qquad 
P_i\,P_j=\de_{ij}P_j \und \sum_i P_i=\unity\ ,
\ee
based on a partition of the identity into orthogonal projectors~$P_i$
associated to the extrema~$\phi_i$, allowing for zero projectors.
This is easily checked,
\be
\begin{aligned}
\prod_k(\Ph{-}\phi_k) &\=
\prod_k(\sum_i\phi_iP_i-\phi_k\unity) \=
\prod_k\sum_i(\phi_i{-}\phi_k)P_i \\
&\= \sum_{i_1,\dots,i_n} \prod_k(\phi_{i_k}{-}\phi_k)P_{i_k} \=
\sum_i P_i \prod_k(\phi_i{-}\phi_k) \= 0\ .
\end{aligned}
\ee
Hence, 
\be
E^{(\th)}[\Ph]
\quad\buildrel{\th\to\infty}\over\longrightarrow\quad
\pi\th\sum_i V(\phi_i)\ \tr\,P_i \ =:\ \th\,E_0
\ee
is degenerate over the infinite-dimensional moduli space
$\ \frac{\textrm{U}(\Fcal)}{\bigotimes_i \textrm{U(rank$P_i$)}}\ $
at~$\th{=}\infty$.
At a finite value of~$\th$ I should expand around $\th{=}\infty$
\cite{Gopakumar:2001yw,Hadasz:2001cn},
\bea
&&E_{(\th)}[\Phi_{\textrm{cl}}] \= \th E_0+E_1+\sfrac{1}{\th}E_2 +\ldots
\qquad\textrm{for}\qquad
\Phi_{\textrm{cl}} \= \Ph + \sfrac{1}{\th} \Phi' +\ldots \ ,\\[12pt]
&&\textrm{with}\qquad
E_1 \= -\sfrac{\pi}{2}\sum_{i,j}\phi_i\phi_j\ \tr\,\bigl\{
[\sfrac{X}{\sqrt{\th}},P_i][\sfrac{X}{\sqrt{\th}},P_j] +
[\sfrac{Y}{\sqrt{\th}},P_i][\sfrac{Y}{\sqrt{\th}},P_j] \bigr\}\ .
\eea

\subsection{Double-well example}

I illustrate the above with a double-well potential
\be
V(\phi) \= \bt\,(\phi^2-1)^2  \qquad\longrightarrow\qquad
V'(\phi) \= 4\bt\,(\phi{+}1)\phi(\phi{-}1)
\ee
and link projectors $P(\phi_i)$ to the three extrema $\ \phi_i=-1,0,+1$,
\be
P(-1) =: P \ ,\qquad P(0) = {\bf 0} \ ,\qquad P(+1) = \unity{-}P \ .
\qquad\qquad \textrm{Hence,}
\ee
\begin{minipage}{200pt}
\epsfig{file=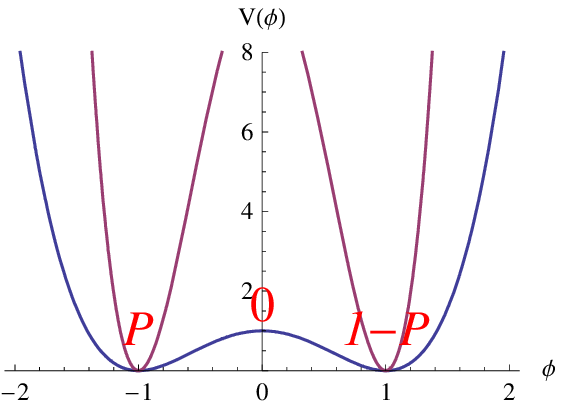}
\end{minipage}
\vskip-130pt
\bea
\phantom{XXXXXXXXXXXXXXX}
&& \Ph\=-1\cdot P +0\cdot{\bf 0} +1\cdot(\unity{-}P)\=\unity-2P \ ,\\[16pt]
&& E_0\= \pi\,\bigl(V(-1)\,P + V(+1)\,(\unity{-}P)\bigr) \= 0 \ ,\\[16pt]
&& E_1 \= -2\pi\,\tr\,\bigl\{
[\sfrac{X}{\sqrt{\th}},P]^2 + [\sfrac{Y}{\sqrt{\th}},P]^2 \bigr\} \\[8pt]
&& \phantom{E_1} \=4\pi\,\tr\,\bigl\{P\ +\ 2|(\unity{-}P)\,\pab P|^2\bigr\}
\quad \text{if \ $\tr\,P<\infty$}\ ,
\eea
where I introduced the abbreviation
\be
\pab\ :=\ [\sfrac{X+\im Y}{\sqrt{2\th}},\cdot\,] \=
[\sfrac{Z}{\sqrt{2\th}},\cdot\,]\ =:\ [a,\cdot\,]
\ee
in the last line. Clearly, BPS saturation $E_1=4\pi\,\tr P$ is reached
for $\pab$-stable projectors, i.e.~when 
$\ a\!: \textrm{im}P \hookrightarrow \textrm{im}P$.
It is easy to see that all further corrections~$\phi'$, $E_2$, etc.\
are of $O(1/\bt)$, whence $\Ph$ and $E_1$ become exact for $\bt\to\infty$
\cite{Klawunn:2006jn}.

\section{Sigma model solitons}

\subsection{Deformed sigma models}

The double-well example is directly related to a sigma model,
since the $\bt\to\infty$ limit nails $\phi$ to the vacuum `manifold'
$\{-1,+1\}$. For a more interesting situation, let me take $\phi$ to be 
complex valued and change the double-well to a Mexican-hat potential, so that
\be
\bt\to\infty \qquad\longrightarrow\qquad \Phi\,\Phi^\+ =\unity\ ,
\ee
i.e.~we have a $\textrm{U}(1)_\th$ sigma model.\footnote{
The $\th$ subscript means that the U($1$)-valued $\phi$ lives on the
Moyal plane or that $\Phi$ acts on $\C\otimes\Fcal$.}
The nonabelian generalization to $\Phi\in\textrm{U}(n)_\th$ is obvious; I just 
have to perform the traces over both $\C^n$ and $\Fcal$, as for example in
\be
E^{(\th)} \= 2\pi\,\tr\ \pa\Phi^\+\pab\Phi \=
2\pi\,\tr\,\bigl| [a,\Phi] \bigr|^2
\qquad\text{with}\qquad[\,a\,,\,a^\+] =\unity\ , 
\ee
where the Heisenberg algebra follows from $\ [Z\,,\,\bar Z]=2\th\unity$.
The configuration space $\textrm{U}(n)_\th$ contains the important
Grassmannian subsectors 
\be
\textrm{Gr}(k,\infty) \= 
\frac{\textrm{U}(n)_\th}{\textrm{U($k$)}\times\textrm{U($\infty{-}k$)}} \= 
\frac{\textrm{U}(\C^n\otimes\Fcal)}{\textrm{U(im$P$)}\times\textrm{U(ker$P$)}}
\qquad\textrm{for}\quad k=\tr\,P,
\ee
which are the spaces of hermitian~$\Phi$ 
based on hermitian rank-$k$ projectors,
\be
\Phi^\+=\Phi \quad\leftrightarrow\quad
\Phi^2=\unity\quad\leftrightarrow\quad
\sfrac12(\unity{-}\Phi) \= P \= P^2 \quad\leftrightarrow\quad
\Phi \= \unity-2P\ .
\ee
In each Grassmannian I can define a topological charge
\be
Q\=\sfrac{1}{8}\,\tr\ \Phi\,[\pa\Phi\,,\pab\Phi]\quad (\=\tr\,P
\quad\textrm{if $\tr\,P$ is finite}\ )\ ,
\ee
and the energy has a lower BPS bound
\cite{Lechtenfeld:2001uq}
\be\label{BPS}
E_\textrm{BPS} \= 8\pi\,|Q| 
\qquad\textrm{attained for}\qquad (\unity{-}P)\,\pab P=0\ .
\ee
Adding a constant to $\Phi$ does not change $Q$, but
interchanging $P$ with $\unity{-}P$ flips its sign.

\subsection{Noncommutative solitons}

In the commutative U($n$) sigma model for $n{>}1$, 
the BPS configurations
\be
\phi=\unity{-}2P \qquad\textrm{with}\qquad (\unity{-}P)\,\pab P=0
\ee
are static solitons, provided the projector $P(x,y)$ is a polynomial 
$n{\times}n$ matrix function of $x$ and~$y$.
A constantly moving single-soliton configuration is generated by a
Lorentz boost,
\be
P(x,y) \ \mapsto\ P\bigl(
\sfrac{x{-}\vel_xt}{\sqrt{1{-}\vec{\vel}^2}},
\sfrac{y{-}\vel_yt}{\sqrt{1{-}\vec{\vel}^2}} \bigr) 
\und E\ \mapsto\ \sfrac{1}{\sqrt{1{-}\vec{\vel}^2}}\,E \ .
\ee
The same is true for the Moyal-deformed theory (via $x\to X$ and $y\to Y$),
but here arises the novel possibility of abelian solitons as I will outline
\cite{Lechtenfeld:2001aw,Lechtenfeld:2001gf}.
In any case, these solitons extremize the action
\be
S_2\=-\sfrac12\int\!\diff{t}\;\pi\th\ \tr\
\eta^{\m\n}\,\pa_\m\Phi^\+\,\pa_\n\Phi
\= \sfrac{\pi\th}2\int_t\,\tr\ A\,\we *\!A
\qquad\textrm{with}\quad A\ :=\ \Phi^\+\,\diff \Phi
\ee
being an auxiliary flat connection.
In a non-integrable model, we cannot expect exact {\it multi-\/}soliton 
solutions. Indeed, numerical studies in the commutative case show
that scattered sigma-model lumps are unstable: they radiate and shrink,
but may live long enough to repel each other and even scatter head-on at 
$90^\circ$.
I can, however, pass to an integrable model (featuring exact multi-solitons)
by adding a WZW-like term~$S_3$ to the action. I will consider the resulting
so-called Ward model~\cite{Ward} from now on.

\section{Ward model solitons}

\subsection{U(n) Ward model}

I begin with the undeformed situation and add to $S_2$ the WZW-like term
\cite{Ward}
\be
S_3 \= -\sfrac13 \int_0^1\int_{\R^{2,1}}\
V\wedge\tr\ \At\we\At\we\At \qquad \textrm{with an extension}\quad
\At(\la{=}\begin{smallmatrix}1\\0\end{smallmatrix}) =
\begin{smallmatrix}A\\0\end{smallmatrix}
\ee
and a Lorentz-breaking one-form $\ V = \diff{x}$.
The equation of motion then changes to
\be\label{Yangeom}
0 \= (\eta^{\m\n}+V_\r\,\ve^{\r\m\n})\,\pa_\m(\phi^\+\,\pa_\n\phi) \=
\pa_x(\phi^\+\,\pa_x\phi) + \pa_{y-t}(\phi^\+\,\pa_{y+t}\phi)
\ee
and remains boost invariant only in the $y$~direction.
Static solitons do not see~$S_3$ and, hence, still have the form
\be
\phi=\unity{-}2P \qquad\text{with the BPS condition}\quad \pab P=P\,\Ga\ ,
\ee
where $\Ga(x,y)$ is an arbitrary $n{\times}n$ matrix function.
Due to the reduced Lorentz invariance, however, 
their boosted cousins may suffer a shape change~\cite{Ward},
\be\label{boosted}
\phi \= (\unity{-}P)+\sfrac{\m}{\mb}P \= \unity-(1{-}\sfrac{\m}{\mb})\,P\ ,
\quad\textrm{with}\qquad
\m\=-\sfrac{\vel_x+\im\sqrt{1{-}\vec{\vel}^2}}{1-\vel_y}\ \in\C\setminus\R
\ee
serving as a rapidity parameter, and with a boosted argument of~$P$. 
For $\vel_x{=}0$ I note that $\im\m$ becomes the proper Lorentz 
contraction factor, and $\phi=\unity{-}2P$ again. 
Clearly, $\m{=}0$ is the static case.
Each polynomial BPS projector~$P$ then yields a soliton with velocity
$(\vel_x,\vel_y)$ and energy $E$ given by (see Figure~1)~\cite{Ward}
\begin{figure}
  \includegraphics[height=.3\textheight]{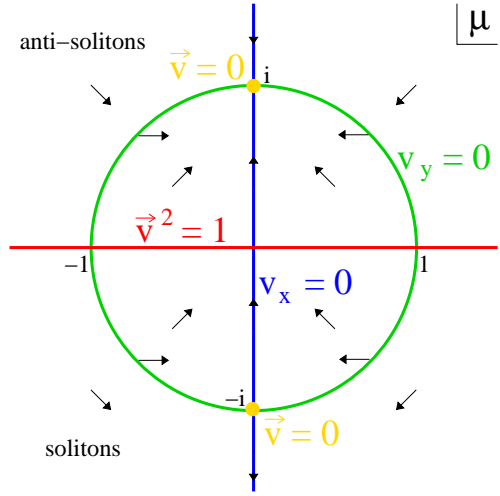}
  \caption{Velocity $\vec\vel$ as function of rapidity $\mu$}
\end{figure}
\be\label{vE} 
(\vel_x,\vel_y) \= -\bigl(
\sfrac{\mu+\bar\mu}{\mu\bar\mu+1}\ ,\ \sfrac{\mu\bar\mu-1}{\mu\bar\mu+1}\bigr)
\und
E \= \sfrac{\sqrt{1-\vec{\vel}^2}}{1-\vel_y^2}\ 8\pi\;\text{rank}\,P
\= \sfrac{1-\vec{\vel}^2}{1-\vel_y^2}\,\sfrac{E_0}{\sqrt{1{-}\vec{\vel}^2}}\ .
\ee

\subsection{Relation with 2+2 self-dual Yang-Mills and 2+1 Bogomolny}

Being integrable, the Ward sigma model should descend from self-dual Yang-Mills
in $2{+}2$ dimensions via the Bogomolny system in $2{+}1$ dimensions
\cite{Ward}.
Their first-order field equations imply the second-order full Yang-Mills and
Yang-Mills-Higgs equations, respectively:
\be
\begin{aligned}
d{=}2{+}2: \qquad{} && 
D^m f_{mn}=0 \qquad\Longleftarrow\qquad f_{mn}=\sfrac12\eps_{mnpq}f^{pq}\\
\downarrow \qquad\qquad && 
\qquad \downarrow \qquad\qquad\qquad\qquad\qquad\quad\downarrow\qquad\quad{} \\
d{=}2{+}1: \qquad{} && 
\begin{smallmatrix} D^\mu f_{\mu\nu}\=h D_\nu h\\
                    D^\mu D_\mu h \=0 \end{smallmatrix}
\qquad\Longleftarrow\qquad f_{\mu\nu}=\eps_{\mu\nu\rho}D^\rho h
\end{aligned}\quad,
\ee
with obvious notation.
A light-cone ansatz~\cite{Yang},
\bea\label{lc1}
&& a_{t-y} \= 0 \qquad\quad\ \ \;\und
a_x-h \= 0 \ ,\\ \label{lc2}
&& a_{t+y} \= \phi^{-1}\,\pa_{t+y} \phi \,\und
a_x+h \= \phi^{-1}\,\pa_x \phi
\eea
with $\ \phi(t,x,y)\in$ {U(n)},
solves two of the three Bogomolny equations. 
The remaining equation precisely yields the (second-order) 
``Yang equation''~(\ref{Yangeom}) for Ward's sigma model,
\be 
\pa_x ( \phi^\+ \pa_x \phi )\ +\ \pa_{y-t} ( \phi^\+ \pa_{y+t} \phi ) \=0\ .
\ee
Therefore, (\ref{BPS}) is a ``second-stage'' BPS bound.

\subsection{Ward multi-solitons}

The Ward model features a Lax pair and a linear system, which may be employed
to construct exact nonabelian commutative multi-soliton configurations. 
The analysis reveals two types of solutions:
First~\cite{Lechtenfeld:2001aw}, 
generically one finds no-scattering configurations
$\ \phi=\prod_k\bigl(\unity-(1{-}\sfrac{\m_k}{\mb_k}P_k\bigr)\ $ 
based on a collection $\ \{(P_k\,,\,\mu_k)\}_{k=1\ldots m}\ $ 
of (non-orthogonal) individually boosted projectors. 
Their energy densities show $m$ lumps moving about 
the $xy$~plane with constant velocities~$\vec{\vel}_{k=1\ldots m}$ and 
oblivious to each other. 
Second~\cite{Lechtenfeld:2001gf}, 
in asymptotically-coinciding-velocity limits 
$\vec{\vel}_k{-}\vec{\vel}_\ell\to0$ for $t\to\pm\infty$,
novel scattering configurations emerge. For instance, 
two lumps `$k$' and `$\ell$', without relative motion at infinity, 
accelerate toward each other and scatter (at rational angles when head-on). 
Alternatively, one finds ring-like time-dependent multi-soliton bound states. 
In the noncommutatively deformed Ward model, such configurations occur even 
in the abelian case, i.e.~for $\textrm{U}(1)_\th$. To understand these,
I must characterize the corresponding projectors.

\subsection{Projectors of infinite and finite rank}

Recall that soliton configurations are constructed from 
hermitian projectors~$P$.
In the undeformed $\tU(n{\ge}2)$ (Ward) sigma model, 
$P$ simply acts on the ``color space''~$\C^n$ and has ``color rank'' $r_c$.
In the deformed $\tU(n{\ge}1)_\th$ (Ward) sigma model, however, $P$ acts 
on the product space $\C^n\otimes\Fcal\simeq\Fcal\oplus\dots\oplus\Fcal$,
with a ``total rank'' $r=r_1+\dots+r_n$ in a color-diagonal basis.
The obvious situation is a smooth deformation $\ P\ \to\ P(X,Y)\ $ 
of a commutative diagonal color-rank-$r_c$ projector 
to an operator-valued diagonal $n{\times}n$ matrix, with a total rank 
$\ r=\infty_1+\ldots+\infty_{r_c}+0+\ldots+0 = r_c\cdot\infty$.
Yet, this is a very special case:
In any block `$i$' I could take $0<r_i<\infty$ and have $r\neq r_c\cdot\infty$,
impeding a smooth $\th\to0$~limit (since $r_c$ is ill-defined)
\cite{Domrin:2004pg}.
The epitome of such a genuinely noncommutative configuration has 
$\ r_i<\infty\quad\forall i\ $, so that the total rank~$r$ is finite. 
In particular, I may now choose $n=1$ and $P$ of finite rank~$r$ inside $\Fcal$.
These ``abelian projectors'' are the most simple and also the most extremely
noncommutative ones, and I will concentrate on them for the rest of this talk.

\section{Abelian multi-solitons}

\subsection{Static abelian solitons}

How do the static $\textrm{U}(1)_\th$ solitons look like?
I know that these are Grassmannian BPS configurations in U($\Fcal$), i.e.\
\cite{Lechtenfeld:2001uq}
\be
\Phi \= \unity{-}2P \qquad\textrm{with}\qquad
(\unity{-}P)\,a\,P\,=\,0\ .
\ee 
Any rank-$r$ hermitian projector in $\Fcal$ decomposes via
\be
P \= |T\> \sfrac{1}{\<T|T\>} \<T| \qquad\textrm{with}\qquad
|T\> \= \bigl( |T_1\>, |T_2\>, \dots, |T_r\> \bigr)\ .
\ee
The collection of ket states $|T_i\>\in\Fcal$ (arranged in a row)
spans im$P$ but is otherwise arbitrary.
The BPS condition above then translates to $|T\>$ as
\be\label{TBPS}
a\,|T\>\,=\, |T\>\,A \qquad\textrm{for some $r{\times}r$ matrix~$A$}\ ,
\ee
i.e.~I am looking for $a$-stable subspaces of dimension~$r$.
Generically I can diagonalize 
\be
A\quad\to\quad\text{diag}(\a_1,\a_2,\dots,\a_r)
\ee
by a basis change inside im$P$.\footnote{
The general case gives Jordan blocks which pose no problem
\cite{Gopakumar:2001yw}.}
Hence, the $|T_i\>$ are eigenkets of~$a$, also known as coherent states
\cite{Gopakumar:2000zd,Lechtenfeld:2001aw},
\be\label{cohsta}
|T\>\=\bigl( |{\a_1}\>, {\a_2}\>, \dots, |{\a_r}\> \bigr)
\qquad\textrm{with}\qquad |{\a_i}\>\ \propto\ \e^{{\a_i}a^\+ }\,|0\>
\quad\textrm{and}\quad a|0\>=0\ .
\ee
The corresponding BPS projector reads
\be
P_r \= \sum_{i,j=1}^r
|{\a_i}\> \,\bigl( \<{\a_.}|{\a_.}\> \bigr)^{-1}_{ij} \<{\a_j}|\ ,
\ee
where the matrix of coherent-state overlaps must be inverted.
For illustration I display the rank-one and rank-two cases
(denoting $|1\>=a^\+|0\>$), 
\bea
P_1 &=& \sfrac{|\a\>\<\a|}{\<\a|\a\>}
\= \e^{-|\a|^2} \e^{\a a^\+} |0\>\<0| \e^{\bar\a a}
\quad\buildrel{\a\to0}\over\longrightarrow\quad |0\>\<0| \ ,\\
P_2 &=& \sfrac{
|\a\>\<\b|\b\>\<\a| + |\b\>\<\a|\a\>\<\b| -
|\a\>\<\a|\b\>\<\b| - |\b\>\<\b|\a\>\<\a| }{
\<\a|\a\>\<\b|\b\> - \<\a|\b\>\<\b|\a\> }
\ \buildrel{\a,\b\to0}\over\longrightarrow\ |0\>\<0|+|1\>\<1|\ .
\eea

\subsection{Noncommutativity in function space: Moyal star product}

In order to develop some intuition for the properties of the noncommutative
solitons and their deviation from the commutative ones, 
I may convert the Moyal-Weyl map~(\ref{MWmap}) and associate my operators
on~$\Fcal$ to ordinary functions on the $xy$~plane (with values in~U($n$)).
For this to be a homomorphism of the operator composition algebra,
I must however deform the pointwise function product to the Moyal star product
\be
\begin{aligned}
(f\sta g)(x,y)&\= f(x,y)\,\exp\,\bigl\{ \sfrac{\im}{2}\th\,
({\ld{\partial}}_x {\rd{\partial}}_y -
 {\ld{\partial}}_y {\rd{\partial}}_x ) \bigr\}\,g(x,y) \\
&\=f\,g\ +\ \sfrac{\im\th}{2}\,(f_x\,g_y{-}f_y\,g_x)
 \ -\ \sfrac{\th^2}{8}\,(f_{xx}\,g_{yy}{-}2f_{xy}\,g_{xy}{+}f_{yy}\,g_{xx})
 \ +\ldots \\[4pt]
&\=f\,g\ +\ \text{total derivatives}\ .
\end{aligned}
\ee
The inverse of (\ref{MWmap}) then maps (usually not Weyl-ordered) operators
$F=F(X,Y)$ to functions $f_\th=F_\sta(x,y)$, where the star indicates
the Moyal product in all compositions.
The most important properties of the deformed product are
\be\label{ncprop}
(f\sta g)\sta h \= f\sta(g\sta h) \quad,\qquad
\smallint\!\!\diff{x}\diff{y}\ f\sta g \= \smallint\!\!\diff{x}\diff{y}\, f\,g
\quad,\qquad x\sta y-y\sta x \= \im\,\th \ ,
\ee
and the key object for building the abelian solitons is the 
operator $\leftrightarrow$ function
\be
|{\a}\>\<{\beta}| \quad\longleftrightarrow\quad
2{\e\!}^{\im\!\kappa} \e^{-\frac12|{\a}-{\beta}|^2}
\e^{-(z-\sqrt{2\th}{\a})(\zb-\sqrt{2\th}{\beta})/\th}
\ \propto\ \e^{-[(x-{x_0})^2+(y-{y_0})^2]/\th}
\ee
with constants $x_0$, $y_0$ and $\kappa$ depending on $\a$ and $\b$.
Clearly, the abelian soliton profile is Gaussian, as is its energy density.
This representation also makes the singular $\th\to0$ limit explicit.

\subsection{Abelian Ward solitons}

Let me now construct the moving solutions (\ref{boosted}) for the abelian case.
To formulate the BPS condition efficiently, 
I pass from the standard coordinates~$(z,\bar z,t)$
to co-moving (or rest-frame) coordinates~$(w,\bar w,s)$ in a linear way
\cite{Lechtenfeld:2001uq,Chu:2005tv}
as depicted in Figure~2.
We also rescale $w$ such that the coordinate change is canonical, or
preserves the noncommutativity relation (\ref{ncdef}) or (\ref{ncprop}),
i.e.~$[w,\bar w]_\sta=2\th\=[W,\bar W]$. 
\begin{figure}
  \includegraphics[height=.3\textheight]{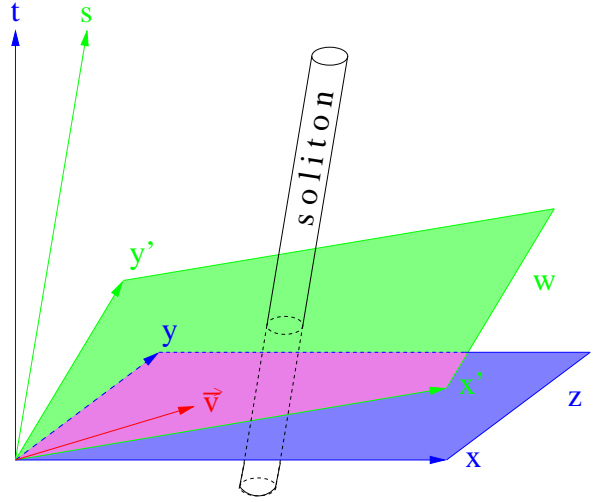}
  \caption{Soliton worldline and co-moving coordinates}
\end{figure}
Normalizing
\be
W \= \sqrt{2\th}\,c \qquad\textrm{such that}\qquad [c,c^\+]=\unity\ ,
\ee
$c$ is connected with $a$ by an ISU(1,1) squeezing (or inhomogeneous
Bogoliubov) transformation~\cite{Lechtenfeld:2001aw},
\be
c \= S(t)\,a\,S(t)^\+ \und |\vel\> \= S(t)\,|0\> 
\qquad\textrm{with}\quad c\,|\vel\>=0\ ,
\ee
where I denote with $|\vel\>$ the co-moving vacuum ket.
The BPS condition them simply reads 
\be
\pa_s P \= 0 \und
(\unity{-}P)\,c\,P \= 0 \qquad\longleftrightarrow\qquad
c\,|T\> \= |T\>\,A
\ee
and is solved by $c$-coherent states (see (\ref{cohsta}))
\cite{Lechtenfeld:2001aw}
\be
|T_i\>\ \propto\ \e^{{\a_i} c^\+ }|\vel\> \= S(t)\,|\a_i\>\ .
\ee
It turns out that the soliton velocity and energy do not depend
on~$\th$, but the $\text{U}_\sta(1)$ soliton configuration is 
singular for~$\th{\to}0$.
The rank-one example reads
\be\label{movsol}
\Phi \= \unity-(1{-}\sfrac{\m}{\mb})\,S(t)|\a\>\<\a|S(t)^\+
\quad\ \ \leftrightarrow\quad\ \
\phi \= 1-(1{-}\sfrac{\m}{\mb})\,2 \e^{-|w(z,t)-\sqrt{2\th}{\a}|^2/\th}
\ee
and represents a squeezed Gaussian lump moving with constant speed.

\subsection{Abelian Ward multi-solitons}

It is now clear how the construction of commutative Ward multi-solitons
mentioned in subsection~3.3 carries over to the noncommutative situation.
A $\text{U}_\sta(1)$ $m$-soliton configuration with rapidities
$(\m_1,\dots,\m_m)$ factorizes as 
\be
\Phi \= \prod_{k=m}^1 \bigl( \unity\ -\ (1{-}\sfrac{\m}{\mb})\,P_k\bigr)
\qquad\textrm{with}\qquad
P_k \= |T^{(k)}\> \sfrac{1}{\<T^{(k)}|T^{(k)}\>} \<T^{(k)}|\ .
\ee
The (rows of) kets
$|T^{(k)}\>=\bigl(|T_i^{(k)}\>\bigr)_{i=1,\dots,r_k}$
for $k=1,\dots,m$ can be found via
\cite{Lechtenfeld:2004qh}
\be
|T_i^{(k)}\> \= 
\biggl\{\prod_{l=1}^{k-1}\Bigl(\unity\ -\ \sfrac{
\mu_{k-l}-\bar\mu_{k-l}}{\mu_{k-l}-\bar\mu_{k}}\ P_{k-l}\Bigr)\biggr\}
\;S_k(t)\,|\a_i^k\>\ ,
\ee
where the factor in braces describes the background of lumps 1 to~$k{-}1$ 
being felt by lump~$k$ when it is added to the system. 
This allows for a recursive construction. 
As an illustration I present the simplest two-soliton example with
$r_1=r_2=1$~\cite{Lechtenfeld:2001aw,Klawunn:2006jn}:
\be\label{twosol}
\Phi^\+ \= \unity\ -\ \sfrac{1}{1-\m|\s|^2} \,\Bigl\{\,
\sfrac{\m_{11}}{\m_1}|{1}\>\<{1}|\ +\
\sfrac{\m_{22}}{\m_2}|{2}\>\<{2}|\ -\
\s\m\sfrac{\m_{21}}{\m_2}|{1}\>\<{2}|\ -\
\bar\s\m\sfrac{\m_{12}}{\m_1}|{2}\>\<{1}| \,\Bigr\}
\ee
\be
\textrm{with}\qquad 
|{k}\>=S_k(t)|\a^k\> \ ,\qquad 
\s=\<{1}|{2}\> \ ,\qquad 
\m_{ij}=\m_i{-}\bar\m_j \ ,\qquad 
\m=\sfrac{\m_{11}\,\m_{22}}{\m_{12}\,\m_{21}}\ .
\ee

\subsection{Abelian soliton scattering ?}

The static soliton moduli space $\C^r$ 
is parametrized by $\{\a_i\}_{i=1,\ldots,r}$
which indicate the positions of the lumps in the Moyal plane. 
The bosonic nature of the lumps leads to identification under permutations, 
$\ \C^r\to\C^r/S_r$. I showed that generic abelian multi-solitons are squeezed 
gaussian lumps, which move with mutually distict velocities $\vec\vel_k$
and do not scatter.
As in the commutative case, more interesting time dependencies arise
in coinciding-velocity limits. The simplest case is the two-soliton
solution~(\ref{twosol}), which for $\mu_{1,2}{\to}-\im$ (static center of mass)
tends to~\cite{Lechtenfeld:2001aw,Klawunn:2006jn}
\be
\Phi=(\unity{-}2P)(\unity{-}2\Pt) \quad\textrm{with}\quad
|T\>=|\ar\> \quad\textrm{and}\quad 
|\Tt\>=|\ar\>-\im {t}\sqrt{\sfrac{2}{\th}}\,(a^\+{-}\bar\ar)|\ar\>\ .
\ee
Taking $\a{=}0$ puts the center of mass at the origin and yields
\be
\phi \= 1\ -\ \sfrac{8{t}}{\th+2{t}^2} \Bigl\{
2\sfrac{z\bar z}{\th}\,{t} -\im(z+\bar z)\Bigr\} \,\e^{-z\bar z/\th}\ ,
\ee
which is revealed not as a scattering configuration but as a ring-shaped 
bound state, which performs a single breath (see Figure~3).
There do not seem to occur exact abelian Ward-model scattering solutions.
\hfill \quad{\lower10pt\hbox{
$\buildrel{\bullet\,\bullet}\over{\buildrel{\angle\ }\over\frown}$} }
\begin{figure}
  \includegraphics[height=.2\textheight]{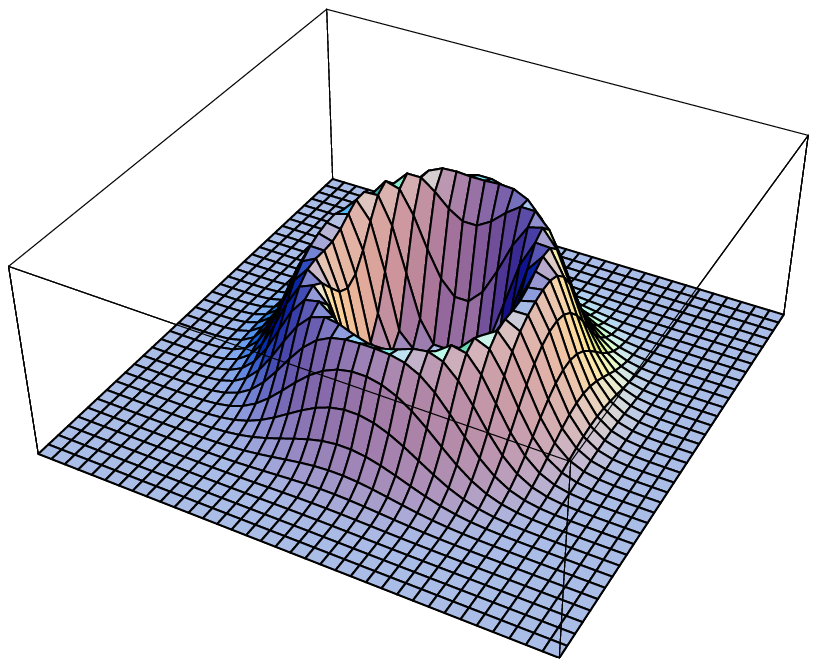} \qquad
  \includegraphics[height=.2\textheight]{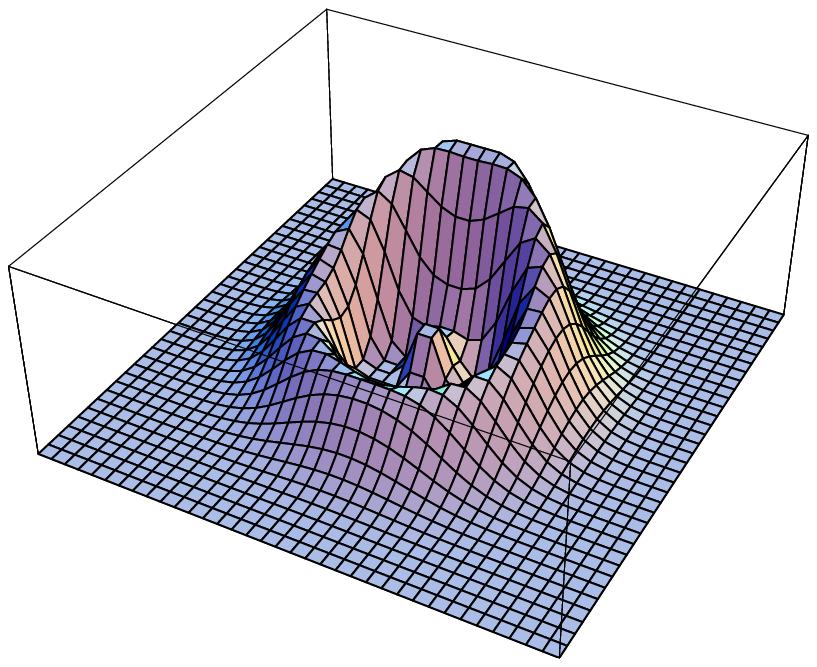} 
\end{figure} 
\begin{figure}
  \includegraphics[height=.2\textheight]{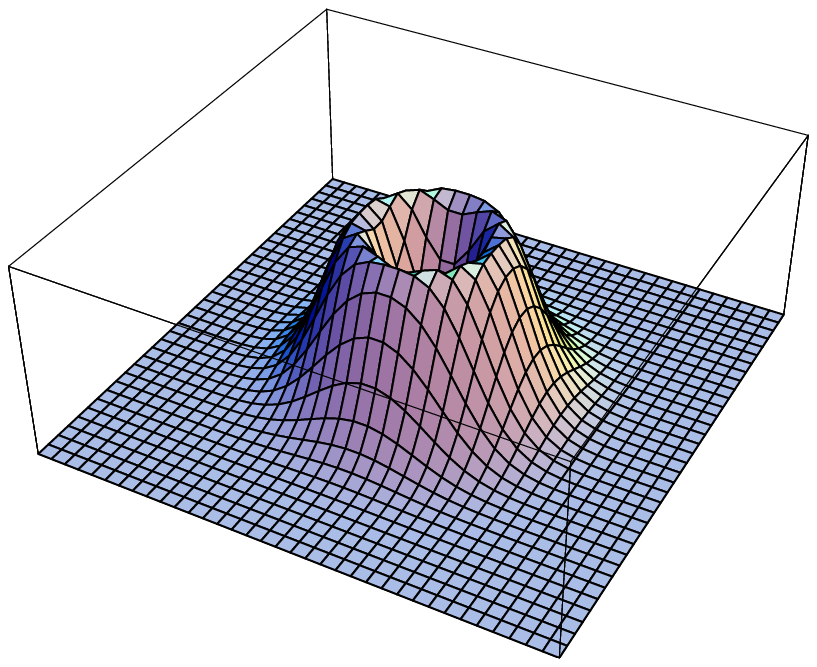} \qquad
  \includegraphics[height=.2\textheight]{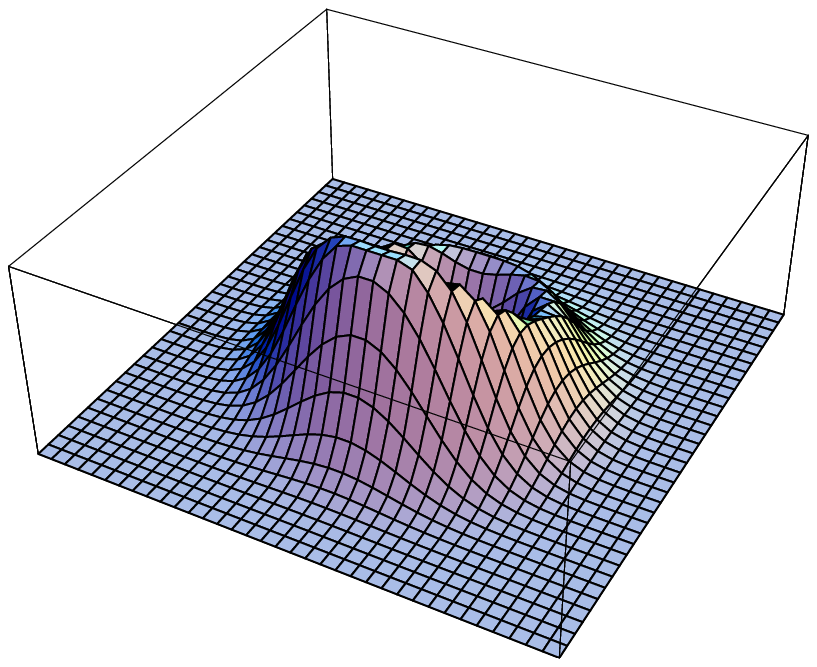}
  \caption{Energy density of abelian two-soliton at 
           $\frac{t}{\sqrt{\th}}\ =\ -5,\,-1.5,\ 0,\ 1.5$}
\end{figure}

\subsection{Stability of static solitons}

An important issue is the stability of my solitons under small perturbations.
The static solitons can be interpreted as solutions in the appropriate
Grassmannian sigma model, where the BPS bound~(\ref{BPS}) guarantees their
absolute stability. When these Grassmannians are imbedded into the general
unitary sigma model, which must be considered for the time-dependent Ward
solitons, topological stability disappears when stepping out of the
Grassmannian. In fact, given any projector inclusion $\Pt\subset P$
(admitting $\Pt=0$), the path~\cite{Domrin:2004pg}
\be
\Phi(s) \= \e\!^{\im s\,(P-\Pt)} (\unity{-}2P)
\= \unity -(1{+}\e\!^{\im s})P -(1{-}\e\!^{\im s})\Pt
\ee
connects $\ \Phi(0)=\Phi=\unity{-}2P\ $ with
$\ \Phi(\pi)=\widetilde\Phi=\unity{-}2\Pt\ $
and thus interpolates between two static solitons in different Grassmannians
inside $\text{U}(\Fcal)$. 
Without knowing any details except for the topological charges $Q$ 
and~$\widetilde Q$, one can compute the energy~\cite{Domrin:2004pg}
\be
\sfrac{1}{8\pi} E(s)
\= \sfrac{Q+\widetilde Q}{2}\,+\,\sfrac{Q-\widetilde Q}{2}\cos s
\= Q\,\cos^2\sfrac{s}{2}\,+\,\widetilde Q\,\sin^2\sfrac{s}{2}\ ,
\ee
which decreases monotonically if $\widetilde Q<Q$.
Therefore, all Ward-model solitons decay to the ``vacuum'' $\Phi=\unity$.
For the diagonal abelian solitons, the instability subspace was shown to
be one-dimensional, and the eigenmode was identified.
In fact, from~(\ref{vE}) I may compare the exact energy
\be
\begin{aligned}
\sfrac{1}{8\pi} E_\text{exact} &\=\
\sfrac{\th}{8}\;\tr \bigl| \dot\Phi(t|\a) \bigr|^2 \ +\
\sfrac14\;\tr \bigl|[a,\Phi(t|\a)]\bigr|^2 \=\
\sfrac{1}{8\pi}\ ( E_\text{kin} \ +\ E_\text{grad} ) \\[4pt]
&\=\ \sfrac{\sqrt{1-\vec{\vel}^2}}{1-\vel_y^2}\
\bigl(\sfrac12\vec{\vel}^2\ +\ 1-\sfrac12\vec{\vel}^2 \bigr) \ \ \approx\ \
\sfrac12 (\vel_x^2 + \vel_y^2)\ +\ 1 - \vel_x^2
\end{aligned}
\ee
of the abelian one-soliton~(\ref{movsol}) at small speed 
to the adiabatic energy 
\be
\sfrac{1}{8\pi} E_\text{adiab} \=\
\sfrac{\th}8\;\tr\bigl|\dot\Phi_{\textrm{static}}(\a{\scriptstyle(t)})\bigr|^2
 \= \sfrac12 (\vel_x^2 + \vel_y^2)\ +\ 1
\ee
in the moduli-space approximation
$\Ph(t|\a)\approx\Ph_{\textrm{static}}(\a{\scriptstyle(t)})$,
after relating $\dot\a$ to~$\vel$~\cite{Klawunn:2006jn}.
The difference is entirely due to the static deformation
$\ 2P\to(1{-}\sfrac{\mu}{\mb})P$:
\be
\sfrac{1}{8\pi} E_\text{deform} \=\
\sfrac14\;\tr \bigl|[a,\Phi(t|\a)|_{S=1}]\bigr|^2 \=\
\sfrac{1-\vec{\vel}^2}{1-\vel_y^2} \ \ \approx\ \ 1 - \vel_x^2\ .
\ee
Therefore, these solitons loose energy by self-accelerating in the 
$x$~direction, i.e.~the Lorentz violation is also the doom for their stability.

\section{Noncommutative sine-Gordon}

\subsection{Dimensional and algebraic reduction}

The Ward model is a gateway to integrable systems in $1{+}1$ dimensions,
by a process of dimensional and algebraic reduction. To obtain solitons
moving on the line, one may either solve the reduced equations of motion
\cite{Lechtenfeld:2004qh},
or else reduce wave solutions~\cite{Bieling:2002is}
 of the U($n$) Ward model along their invariance direction, say the $x$~axis. 
Specifying to the commutative U(2)~case and introducing the light-cone 
coordinates $u=\sfrac12(t{+}y)$ and $v=\sfrac12(t{-}y)$,
I first perform a (twisted) dimensional reduction by dictating 
a particular $x$~dependence for the solution $\phi(u,v,x)\in\textrm{U}(2)$
\cite{Lechtenfeld:2004qh}, 
\be
\phi(u,v,x) \=
\cE\e^{\im\a\,x\,\se}\;g(u,v)\,\e^{-\im\a\,x\,\se}\,\cE^\+
\qquad\text{for}\quad g(u,v)\in \textrm{U(2)}\ ,
\ee
where $\cE$ is a suitable constant $2{\times}2$ matrix, 
$\sigma_a$ with $a=1,2,3$ are the Pauli matrices, and
$\a\in\R$ is a parameter of the reduction.
The Yang equation of motion~(\ref{Yangeom}) for $\phi$ then turns into
\be\label{geom}
\pa_v ( g^\+ \pa_u g )\ +\ \a^2 (\se g^\+ \se g - g^\+ \se g\,\se) \=0\ .
\ee
As a second step, I algebraically restrict $g(u,v)$ to U(1) via
\be\label{sg} 
g\=\e^{\frac{\im}{2}\sd\,\vp}
\qquad\longrightarrow\qquad
\pa_v \pa_u \vp\ +\ 4\a^2 \sin\vp \= 0
\ee
and discover the sine-Gordon equation for the angular field~$\vp(t,y)$.

\subsection{Time-space deformation}

For Moyal-type noncommutativity to survive the reduction to $1{+}1$
dimensions, I must switch from my space-space deformation to a time-space one
\cite{Chu:2005tv},
\be
[t,y]_\sta\=\im\th\qquad\leftrightarrow\qquad [u,v]_\sta\=-\sfrac{\im}{2}\th\ ,
\ee
whereupon the $g$~equation of motion (\ref{geom}) picks up Moyal stars,
\be\label{ncgeom}
\pa_v(g^\+\sta\pa_u g)\ +\ \a^2(\se g^\+\sta\se g\ -\ g^\+\se\sta g\,\se)\=0\ .
\ee
In this section I will stick with the star-product formulation.
For the algebraic reduction, a restriction to U(1) is no longer appropriate,
since in noncommutative U($n$)~Yang-Mills theory the overall U(1)~phase 
no longer decouples, and one should keep its degree of freedom besides
reducing SU(2) to U(1). With this important insight~\cite{Lechtenfeld:2004qh},
my ansatz becomes
\be
g\=\es^{\frac{\im}{2}{\unity}\,\r}\sta\es^{\frac{\im}{2}\sd\,\vp}
\quad\in\ \textrm{U}_\sta(1)\times\textrm{U}_\sta(1)
\ee
featuring {\it two\/} angular fields $\vp$ and $\r$.
{}From (\ref{ncgeom}) I get their equations of motion,
\be\label{ncsg}
\begin{aligned}
\pa_v\bigl(\!\es^{-\frac{\im}{2}\vp}\sta\pa_u\es^{\frac{\im}{2}\vp} \bigr)\ +\
2\im\a^2\sin_\sta\!\vp &\=
-\pa_v\bigl[\es^{-\frac{\im}{2}\vp}\sta R\sta\es^{\frac{\im}{2}\vp} \bigr]\ ,
\\
\pa_v\bigl(\!\es^{\frac{\im}{2}\vp}\sta\pa_u\es^{-\frac{\im}{2}\vp} \bigr)\ -\
2\im\a^2\sin_\sta\!\vp &\=
-\pa_v\bigl[\es^{\frac{\im}{2}\vp}\sta R\sta\es^{-\frac{\im}{2}\vp} \bigr]\ ,
\end{aligned}
\ee
where $\ R=\es^{-\frac{\im}{2}\r}\sta\pa_u\es^{\frac{\im}{2}\r}\ $
carries the second scalar field $\r$.
The sum of these looses the $\a$~dependence and simplifies to
$\ \pa_v\pa_u\r=0\ $ in the $\th{\to}0$ limit while the difference
converges to the commutative sine-Gordon equation~(\ref{sg}).
I name (\ref{ncsg}) the ``noncommutative sine-Gordon'' equations
\cite{Lechtenfeld:2004qh}.

\subsection{Noncommutative sine-Gordon kinks}

The equations~(\ref{ncsg}) have deformed multi-kink solutions,
which can be constructed via a linear system.
Since from the $2{+}1$~perspective the waves must move in the $y$~direction, 
the rapidities $\ \m=\im p\in\im\R$ are purely imaginary (see (\ref{boosted}))
 and, putting $x{=}0$, the co-moving coordinate~$w$ simplifies to
\be
w \= \bar{\m}u+\sfrac{1}{\bar{\m}}v = -\im (pu+\sfrac{1}{p}v)
\= -\im \sfrac{y-\vel t}{\sqrt{1-\vel^2}}\ =:\ -\im \h
\qquad\textrm{with}\quad \h\in\R\ .
\ee
The Ward-model one-wave solution descends to the one-kink configuration
\cite{Lechtenfeld:2004qh}
\be
g \= \cE^\+\sd(\unity{-}2P)\,\cE \qquad\textrm{with projector}\qquad
P \= T\sta(T^\+{\sta}T)_\sta^{-1}\sta T^\+\ ,
\ee
where the $1{\times}2$ matrix function $T(\h)$ is subject to 
the BPS condition
\be
(\pa_\h + \a\,\sd)\,T \= T\sta A
\ee 
with an arbitrary function $A(u,v)$, which I put to zero.
Conveniently fixing unimportant integration constants, the solution reads
\cite{Lechtenfeld:2004qh}
\bea
&&T \= \biggl(\begin{matrix}
\e^{-\a\h}\  \\[8pt] \im\e^{\a\h} \end{matrix}\biggr)
\qquad\longrightarrow\qquad
P \= \frac{1}{2\,\ch 2\a\h} \biggl(\begin{matrix}
\e^{-2\a\h}\!\! & -\im \\[8pt] \im & \!\!\e^{+2\a\h} \end{matrix}\biggr)
\\[12pt] 
\longrightarrow\ &&g \= \cE^\+
\Biggl(\begin{matrix} \tah 2\a\h & \frac{\im}{\ch 2\a\h}\\[8pt]
\frac{\im}{\ch 2\a\h} & \tah 2\a\h \end{matrix}\Biggr)\,\cE
\ \buildrel{!}\over{\=}\
\Biggl(\begin{matrix}
\es^{\frac{\im}{2}\,\r}\sta\es^{\frac{\im}{2}\,\vp}\!\! & 0 \\[8pt]
0 & \!\!\es^{\frac{\im}{2}\,\r}\sta\es^{-\frac{\im}{2}\,\vp}
\end{matrix}\Biggr)\ .
\phantom{XXX}
\eea
The last equation reveals that 
$\ \cE=\sfrac{1}{\sqrt{2}}\bigl(\begin{smallmatrix}
1 & -1 \\ 1 & \phantom{-}1 \end{smallmatrix}\bigr)\ $
and 
\be
\es^{\frac{\im}{2}\,\r}\sta\es^{\pm\frac{\im}{2}\,\vp} \=
\tah 2\a\h\ \pm\ \sfrac{\im}{\ch 2\a\h}\ ,
\ee
which implies that $\ \r=0\ $ and 
\be
\cos_\sta\sfrac{\vp}{2}=\tah 2\a\h \quad\longrightarrow\quad
\tan_\sta\sfrac{\vp}{4}=\e^{-2\a\h} \quad\longrightarrow\quad
\vp = 4\,\arctan \e^{-2\a\h}\ ,
\ee
which is the standard (undeformed) sine-Gordon kink with velocity
$\vel=\sfrac{1-p^2}{1+p^2}$.
Actually, this was clear beforehand, since a soliton in $1{+}1$ dimensions
can only depend on the single co-moving combination $\h(u,v)$
of the light-cone coordinates, which trivializes the star product
and effectively puts $\th{=}0$ and $\r{=}0$.
In contrast, breather and two-soliton solutions do get deformed however,
according to
\be
[{\h_i}\,,{\h_k}]_\sta\=-
\im\th\,(\vel_i{-}\vel_k)\big/
\sqrt{(1{-}{\scriptstyle{\vel_i^2}})(1{-}{\scriptstyle{\vel_k^2}})}\ .
\ee

\section{Supersymmetric extensions}

Remembering the self-dual Yang-Mills ancestor of the Ward model,
it is straightforward to supersymmetrize its noncommutative generalization
\cite{Lechtenfeld:2007we}.
The maximal $\Ncal{=}4$ extension of noncommutative self-dual Yang-Mills 
reduces to a Moyal-deformed\footnote{
As before, I only deform the bosonic spatial coordinates,
and pass to the operator formulation.}
$2\Ncal{=}8$ super Bogomolny system \
$(F_{\a\b},\ \chi^i_\a,\ \phi^{[ij]},\
 \tilde\chi^{[ijk]}_\a,\ G_{\a\b}^{[ijkl]})$ \
subject to
\bea
&&F_{\a\b}+D_{\a\b}H \=0
\qquad\qquad\qquad\qquad\qquad\text{with}\quad
D_{\a\b} \= \pa_{\a\b} + [ A_{\a\b}\,,\,\cdot\, ]\ ,
\nonumber\\
&&D_{\a\b}\,\chi^{i\b} + \ve_{\a\b}\,[H ,\, \chi^{i\b} ]\=0\ ,
\nonumber\\
&&D_{\a\b}\,D^{\a\b}\phi^{ij} + 2[H ,\, [H,\phi^{ij}]]+
2\{\chi^{i\a},\,\chi^j_{\a}\}\=0\ ,
\\
&&D_{\a\b}\,\tilde{\chi}^{\b[ijk]}-\ve_{\a\b}\,[H,\,
\tilde{\chi}^{\b[ijk]} ]-6[\chi_{\a}^{[i},\ \phi^{jk]} ]\=0\ ,
\nonumber\\
&&D_{\a}^{\ \g} G_{\g\b}^{[ijkl]}{+}
[H,G_{\a\b}^{[ijkl]}]{+}
12\{\chi_{\a}^{[i},\tilde{\chi}_{\b}^{jkl]}\}{-}
18[\phi^{[ij},D_{\a\b}\phi^{{kl}]}]{-}
18\ve_{\a\b} [\phi^{[ij},[\phi^{{kl}]},H ]]=0\ ,
\nonumber
\eea
where $\a,\b=1,2$ are SO(2,1) spinorial 
and $i,j,k,l=1,\dots,\Ncal$ are R-symmetry indices.
The superspace
\ $\R^{2,1|4\Ncal} \ni (x^{\a\b},\vt^{i\a},\h_i^\a)$ \
contains a chiral subspace
\ $\R^{2,1|2\Ncal} \ni (x^{\a\b}{-}\vt^{i\a}\h_i^\b,\h_i^\a)
\equiv(y^{\a\b},\h_i^\a)$,
in which I introduce the Higgs superfield $\Hcal$
plus chiral superfield potentials
$(\Acal_{\a\b},\ \Acal^i_{\a})$
and their field strengths
$(F_{\a\b},\ \chi^i_\a,\ \p^{ij})$
with leading component fields
$(F_{\a\b},\ \chi^i_\a,\ \p^{ij})$.
I extend the light-cone ansatz~(\ref{lc1})--(\ref{lc2}) to
\cite{Lechtenfeld:2007we}
\bea
\Acal_{12}-\Hcal&=& 0 \und\qquad\,
\Acal_{22}\=\Phi^{-1}\pa_{t+y}\Phi\ ,\\
\Acal_{11}&=& 0 \und
\Acal_{12}+\Hcal\=\Phi^{-1}\pa_{x}\Phi\ ,\\[4pt]
\Acal^i_1&=& 0 \und\qquad\quad
\Acal^i_2\=\Phi^{-1}\pa^i_2\Phi\ ,
\eea
and the remaining (superspace) Bogomolny equations yield equations of motion
for the operator-valued U($n$) Yang prepotential chiral superfield \ 
$\Phi(y,\h)$~\cite{Lechtenfeld:2007we}:
\bea
\pa_x(\Phi^{-1}\pa_x\Phi) + \pa_{y-t}(\Phi^{-1}\pa_{y+t}\Phi)\=0\ ,\\[2pt]
\pa_1^i(\Phi^{-1}\pa_x\Phi) + \pa_{y-t}(\Phi^{-1}\pa_2^i\Phi)\=0\ ,\\[2pt]
\pa_1^i(\Phi^{-1}\pa_{t+y}\Phi) - \pa_x(\Phi^{-1}\pa_2^i\Phi)\=0\ ,\\
\pa_1^i(\Phi^{-1}\pa_2^j\Phi) + \pa_1^j(\Phi^{-1}\pa_2^i\Phi)\=0\ .
\eea
The construction of solitons then mimics the bosonic construction but 
takes place in the chiral superspace. The familiar co-moving coordinates
get extended by a fermionic combination, $\h_i=\h_i^1+\bar\mu\h_i^2$.
One finds that the supersymmetric solitons are essentially a
Grassmann-algebra lift of the bosonic ones.
Abelian solitons, for instance, are still based on coherent states, but
with parameters~\cite{Lechtenfeld:2007we} 
\be
\ar(\h) = \ar^0 + \h_i \ar^{[i]} + \h_i\h_j \ar^{[ij]} +
\h_i\h_j\h_k \ar^{[ijk]} + \h_i\h_j\h_k\h_l \ar^{[ijkl]}\ .
\ee
Such nilpotent additions to the soliton profiles and energy densities
however do not qualitatively alter the multi-soliton dynamics
\cite{Gutschwager:2007ev}.
A more promising option includes non-{\it anti\/}commutative deformations,
\be
\{\h_i^{\a}\,,\,\h_j^{\b}\}\=C_{ij}^{\a\b}
\ee
replacing the Grassmann by a Clifford algebra, which should break
half of the supersymmetry and might yield solitonic spin degrees of freedom.


\begin{theacknowledgments}
I am grateful to Abel Camacho, Claus L\"ammerzahl and Alfredo Macias
for having organized a wonderfully stimulating meeting.
\end{theacknowledgments}

\newpage


\bibliographystyle{aipprocl} 


\end{document}